\documentclass[10pt]{iopart}
\usepackage{iopams}
\usepackage[svgnames]{xcolor}
\usepackage{xstring}
\usepackage{graphicx}
\usepackage{subcaption}
\usepackage{soul}
\usepackage{lineno}
\usepackage{orcidlink}
\usepackage{url}
\usepackage{hyperref}
\usepackage{doi}

\hypersetup{
    colorlinks=true,
    urlcolor=blue,
    citecolor=blue
}
\newcommand{\corrections}[1]{#1}
\newcommand{\correctionstwo}[1]{#1} 
\begin{document}

\title[Four years of successful operation of the AdV+ IMC instrumented baffle]{The Instrumented Baffle at the Input Mode Cleaner of Advanced Virgo Plus: Four years of successful operation as a monitor of stray light} 

\author{D. Nanadoumgar-Lacroze$^1$\footnote{Author to whom the correspondence should be addressed.}\footnote{This work is part of the doctoral thesis of D.N.L. within the framework of the Doctoral Program in Physics at the Autonomous University of Barcelona.}\orcidlink{0009-0009-7255-8111}, M. Seglar-Arroyo$^1$\orcidlink{0000-0001-8654-409X}, M. Martinez$^{1,2}$\orcidlink{0000-0002-3135-945X}, Ll. M. Mir$^1$\orcidlink{0000-0002-4276-715X}, O. Ballester$^1$\orcidlink{0000-0002-7126-5300}}

\address{$^1$Institut de Física d'Altes Energies (IFAE), The Barcelona Institute of Science and Technology, Campus UAB, E-08193 Bellaterra (Barcelona), Spain\\
$^2$Catalan Institution for Research and Advanced Studies (ICREA), E-08010 Barcelona, Spain}
\ead{dnala@ifae.es}

\vspace{10pt}
%\begin{indented}
%\item[]\today
%\end{indented}

\begin{abstract}

The suspended end mirror of the Input Mode Cleaner cavity in the Advanced Virgo Plus interferometer was equipped with an instrumented baffle in Spring 2021, serving as a demonstrator of the technology in preparation for the installation of large instrumented baffles in the main arms of the interferometer. This baffle includes tens of sensors positioned near the mirror to enable monitoring of stray light within the cavity. In this contribution, we assess the stability and performance of the instrument after four years of operation. After introducing the main characteristics of the baffle, we study the distribution of stray light and show that the instrumented baffle can be used to monitor laser stability and alignment within the cavity. Finally, we perform a noise level assessment with data from the final stages of the O4b commissioning and conclude that the baffle does not introduce any additional disturbance to the normal operation of the interferometer. 

\end{abstract}

%\tableofcontents

%
% Uncomment for keywords
%\vspace{2pc}
%\noindent{\it Keywords}: XXXXXX, YYYYYYYY, ZZZZZZZZZ
%
% Uncomment for Submitted to journal title message
%\submitto{\JPA}
%
% Uncomment if a separate title page is required
\maketitle
% 
% For two-column output uncomment the next line and choose [10pt] rather than [12pt] in the \documentclass declaration
\ioptwocol

\section{Introduction}

The Virgo detector is a ground-based Michelson interferometer with three-kilometre-long arms equipped with Fabry-Pérot cavities. Its current version, Advanced Virgo Plus (AdV+), results from the successive upgrades of the Virgo~\cite{Accadia2012} and Advanced Virgo detectors~\cite{Acernese2014}. The Phase I of AdV+, conducted between the third and fourth observing runs (O3 and O4), aimed to reduce the impact of quantum noise on Virgo sensitivity~\cite{Acernese2023_future}. AdV+ joined the O4 observing run on April 10, 2024, coinciding with the start of the second period of the run, O4b, until the end of O4 on November 18, 2025~\cite{AdV+_O4}. A six-month extension of O4 (nominated IR1) is now scheduled for fall 2026. 
 
Enhancing detector sensitivity and performance requires updates to the instrument's control and operational procedures, which often constitute the primary focus of commissioning periods. This motivates the development of refined monitoring devices to streamline operation and commissioning activities. These upgrades are driven by the dual objective of increasing the number of gravitational-wave detections and improving data quality. This is accomplished by addressing the primary fundamental noise sources, i.e. quantum and thermal noise, followed by the mitigation of technical noise sources, e.g. stray light and control system noise.

In the context of laser interferometers, stray light is an important contributor to the technical noise budget. \corrections{Stray light originates from laser light that is scattered or reflected out of the nominal optical path and subsequently recombines with the main beam. The resulting stray-light noise arises from the phase difference accumulated between the scattered and main beams. The coupling of stray light to the interferometer output is further modulated by the motion of the scattering surfaces which convert mechanical motion into phase noise in the detector signal. The mirrors are suspended to reduce the direct coupling of seismic disturbances to the cavity length, thereby preserving the cavity resonance. Nevertheless, residual motion of the suspended optics and alignment fluctuations modulate the phase of light scattered from the high-reflectivity surfaces of the mirrors, allowing stray light to couple back into the interferometer signal~\cite{Accadia2012,Acernese2014,Acernese2023_future}.}

Since the first observing run, complementary approaches have been implemented to mitigate stray light~\cite{Acernese2014}. During the first three observing runs, baffles, low-reflective and low-scattering devices designed to dampen stray light, were positioned in the optical cavities of the interferometer, either within the vacuum tubes, vacuum tower walls, or suspended around the mirrors~\cite{Acernese2014}. By absorbing the light scattered off the cavities walls and back-scattered from the mirrors before it interfere with the beam, baffles play an important role in reducing scattered light. 

Although baffles suit their intended purpose as mitigators of laser beam scattering in the cavities, they do not allow for the control or monitoring of the scattered light. To address this point, \corrections{the installation of two large instrumented baffles in the main Fabry-Pérot cavities close to the North and West input mirrors of Virgo has been planned in preparation for the O5 observing run~\cite{Acernese2019, Andres2023, Ballester2022}. They were successfully installed in June 2026.} This technology aims to eventually enhance the interferometer sensitivity by providing us with an active monitoring device, sensitive to cavity misalignments, defects of the test masses, thermal effects in the mirrors modifying the optical configuration, or the interaction of the laser beam with dust particles potentially modifying the stray light distribution inside the cavities. They will improve our knowledge of the interferometer for future upgrades and next-generation detectors. As a pathfinder for the instrumented baffle technology, a prototype was installed in April 2021 in the Input Mode Cleaner (IMC) cavity following Phase I upgrades of AdV+~\cite{Acernese2023_status}. \corrections{Placed in an auxiliary cavity, this prototype is primarily intended to demonstrate that it operates in a non-perturbative manner. Consequently, its potential as a monitoring device is limited in comparison to the second-generation instrumented baffles deployed in Virgo’s main optical cavities.}

\begin{figure}[htb]
    \centering
    \includegraphics[width=\linewidth]{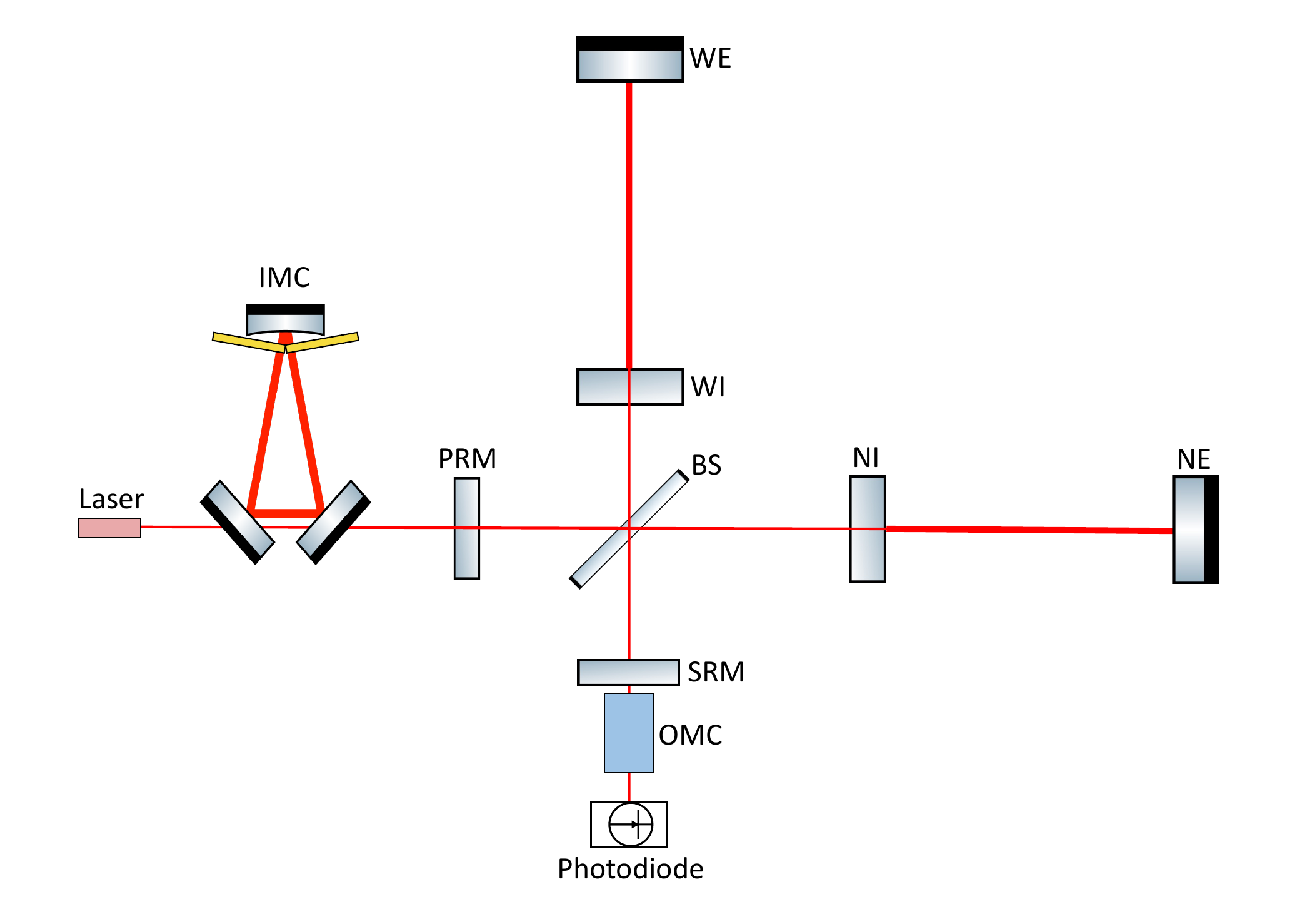}
    \caption{Simplified, not to scale, optical layout of AdV+. Each arm is
formed by an input mirror (North input (NI) or West input (WI)) and an end mirror (North end (NE) or West end (WE)). The IMC cavity is located between the laser source and the Power Recycling Mirror (PRM), prior to the beam splitter (BS). The instrumented baffle is shown in yellow.}
    \label{fig:AdVirgoSchema}
\end{figure}

As shown in Figure~\ref{fig:AdVirgoSchema}, the IMC is a triangular optical cavity prior to the beam splitter and main arms formed by two flat mirrors and a mirror with positive curvature, with a total length of $143.424 \rm ~m$ and a finesse of $\mathcal{O}(10^3)$, matching Virgo's initial design. The primary purpose of the cavity is modal and frequency filtering in vacuum, achieved by stabilising the laser frequency and geometrically cleaning the beam. By reducing amplitude and beam-pointing fluctuations, the IMC significantly enhances the sensitivity of the interferometer~\cite{Acernese2014}.

The IMC baffle is equipped with silicon photodiodes that enable dynamic mapping and direct monitoring of low-angle scattered light, higher-order modes, and contamination or deterioration of mirror surfaces. The illumination of the sensors depends on their geometric distribution relative to the laser beam, as dictated by the triangular configuration of the cavity~\cite{RomeroRodriguez2021}. This fine-tuned instrumentation can facilitate the identification of laser alignment issues and the correlation of observed changes in cavity performance with their underlying causes, providing critical insights into the evolution of observational conditions. The first results were analysed after installation and complied with the expectation from the simulations obtained during the development phase~\cite{Ballester2022, RomeroRodriguez2021}. The technical details of the baffle such as its design, calibration, testing and installation are collected in Ref.~\cite{Andres2023}.

In this paper, we assess the long-term performance of the IMC instrumented baffle and describe how the new information on the IMC cavity complements existing monitoring channels in the interferometer. In Section~\ref{section:Status_Virgo}, we briefly describe the status of the AdV+ detector during the years of IMC instrumented baffle operation. In Section~\ref{section:Baffle_design_DAQ}, we summarise the baffle design and the data acquisition system (DAQ). Section~\ref{section:Stray_light} is devoted to describing the spatial and temporal distribution of the observed stray light in the IMC cavity, and Section~\ref{section:Connection_ITF} connects the baffle, as an active monitoring device, to several aspects relevant for interferometer operations. In Section~\ref{section:Noise_level}, we assess the overall noise level and describe the noise hunting campaign conducted before O4b. 
We conclude by discussing the results and implications for the instrumented baffles in AdV+ Phase II and for the next generation of gravitational-wave interferometers.

\section{Status of the AdV+ detector during data taking}
\label{section:Status_Virgo}

\begin{figure}[htb]
    \centering
    \includegraphics[width=\linewidth]{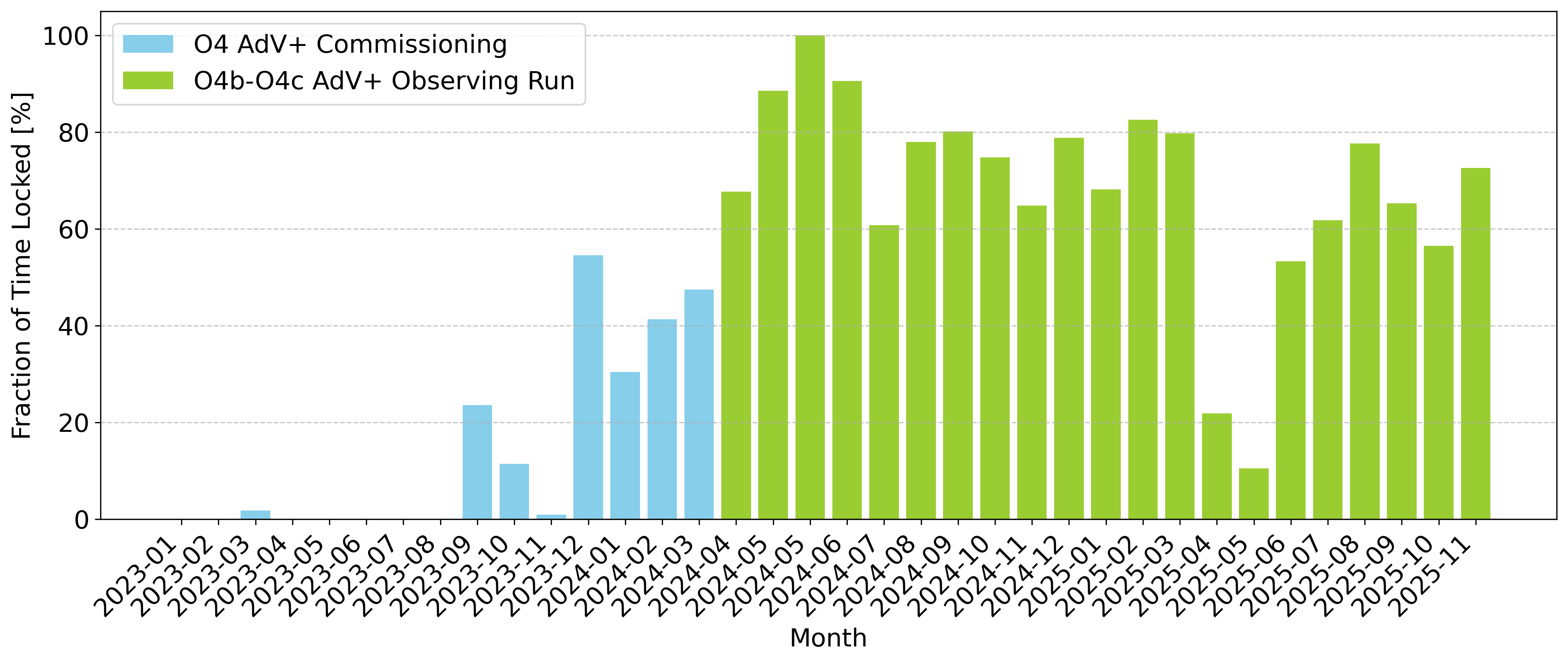}
    \caption{
    Fraction of time during which AdV+ was in a \textit{locked} state (as defined in the text) over the studied period of IMC instrumented baffle operation, from April 2021 to November 2025. The plot differentiates between O4 AdV+ commissioning time and AdV+ participation in the O4 observing run \corrections{(including O4b and O4c, separated by a 10-week commissioning break between April and June 2025)~\cite{AdV+_O4}.}}
    \label{fig:VirgoStatusTimeline}
\end{figure}

The instrumented baffle was installed in April 2021, during the first stage of the interferometer commissioning as part of the AdV+ Phase I upgrade plan. After the instrumented baffle installation, the AdV+ interferometer continued the commissioning activity until April 2024. 

Figure~\ref{fig:VirgoStatusTimeline} illustrates the different Virgo operational phases together with the fraction of time during which the interferometer has been locked since April 2021. For simplicity, we define the interferometer as \textit{locked} when it is in \textit{Low Noise} state, \corrections{with the interferometer maintained in resonance and the noise sufficiently suppressed to enable the highest sensitivity for gravitational wave detection} \footnote{\corrections{The Virgo channel \texttt{V1:META\_ITF\_LOCK\_index} indicates the ITF lock index.} \textit{Low Noise} state is defined as \texttt{V1:META\_ITF\_LOCK\_index} $> 135$.}. \corrections{Although no dedicated Virgo channel directly reports the IMC lock status, the IMC must be locked before lock acquisition can begin for the main arms. Therefore, the interferometer locked fraction provides a conservative lower limit on the IMC locked fraction. Moreover, the interferometer lock status is the most relevant metric for interpreting the baffle measurements, as it identifies periods during which the detector operates in its nominal optical configuration, with the circulating power and beam alignment representative of science-mode operation. An indirect estimate based on the circulating power in the IMC indicates that the IMC remained locked for more than about 95\% of the time throughout the period considered in this work, except during a few periods (e.g., May 2021, April 2022, and November 2023), when commissioning activities temporarily reduced its availability.}

During the O4 commissioning phase, significant efforts were devoted to improving the IMC's operational stability and achieving robust operation at its nominal resonant working point. These activities included changes to the injected laser power and lock acquisition procedure, both of which are expected to influence the scattered-light levels measured by the baffle sensor network. In addition to stabilising the laser frequency~\cite{Acernese2014}, the IMC acts as a filter by suppressing residual higher-order spatial modes and reducing the amount of stray light propagating through the interferometer. Optical simulations presented in Ref.~\cite{RomeroRodriguez2021} further show that both continuous misalignments and transient disturbances can displace the beam within the IMC cavity, increasing optical scattering and, consequently, the amount of light intercepted by the surrounding baffles. \corrections{We refer to Ref.~\cite{AdV+_O4} for further details about the Virgo status during O4.}

\section{Baffle design and data acquisition system}
\label{section:Baffle_design_DAQ}
\begin{figure}[htb]
    \centering
    \includegraphics[width=0.8\linewidth]{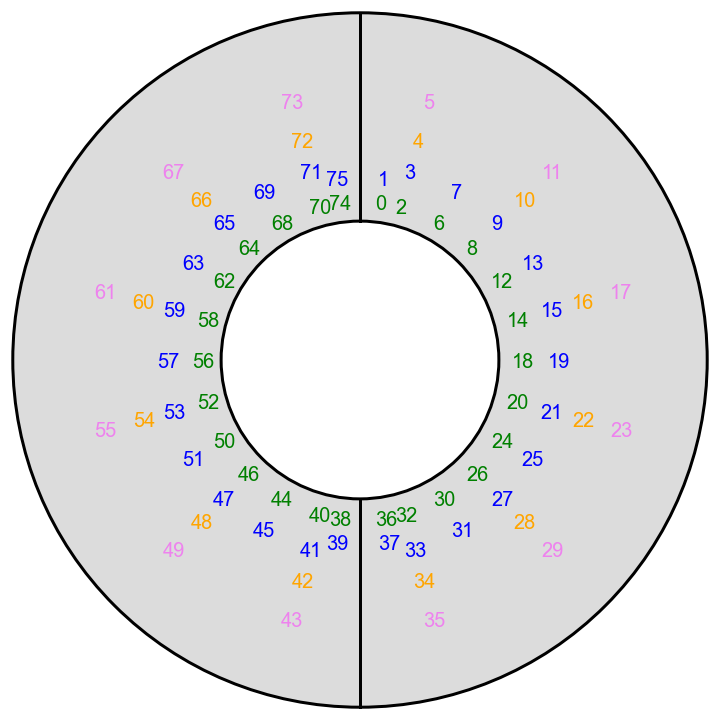}
    \caption{
    Two-dimensional diagram of the instrumented baffle design with labelled sensors, where the colours correspond to the distinct rings and the numbers indicate the sensor indices. \corrections{The rings are numbered in order of increasing radius, with ring 1 shown in green, ring 2 in blue, ring 3 in orange, and ring 4 in pink.}
    }
    \label{fig:BaffleSchema}
\end{figure}

The instrumented baffle is constructed from low-reflective, low-scattering stainless steel and is equipped with $76$ silicon photodiodes arranged in four concentric rings, along with $16$ temperature sensors distributed around the mirror. The sensor placement was motivated by stray light simulations presented in Ref.~\cite{RomeroRodriguez2021}. As illustrated in Figure~\ref{fig:BaffleSchema}, the two innermost rings each contain $26$ sensors, while the two outermost rings each have $12$ sensors, reflecting the expectation that low angles are more illuminated by stray light when the laser beam is properly aligned. \correctionstwo{The baffle is split into two halves, each tilted by 9° with respect to the mirror plane, in order to minimise back-reflections toward the laser beam.} The baffle has inner and outer radii of $7$ cm and $17.5$~cm respectively, with the first ring positioned at $8.1$ cm, the second one at $9.8$~cm, the third one at $11.5$ cm, and the outer ring at $13.5$ cm, as illustrated by Figure~\ref{fig:BaffleSchema}.
Both the baffle and the photodiodes are coated with anti-reflective material, and edges are carefully shaped with small radii of curvature to limit scattering.

The DAQ system records $76$ Analog-to-Digital Converter (ADC) channels readings, 16 temperature readings, the voltage and current supplied to the device. The system operates at a reduced $\rm 2~Hz$ sampling rate (baffles in the input mirrors operate at a 1~kHz sampling rate) for photosensors and $\rm1~Hz$ for temperature sensors, enabling effective monitoring but limiting high-frequency analyses. It relies on 16 ADCs, 8 ADCs per half-baffle. The DAQ board is controlled by a software server, which manages the DAQ boards as well as the power supplies, allowing us to oversee the baffle operation and data storage.

The photodiodes exhibit good linearity, with a sensor-to-sensor variation of $3\%$, and the electronics reach saturation at $\rm15~mW$~\cite{Andres2023}. Scattered light is assessed by converting ADC counts into power using a conversion factor of $\rm 4.6~\mu W/ADC$, previously determined in the laboratory with an uncertainty of $5\%$, while also accounting for sensor-to-sensor variations.

The baffle operates in ultra-high vacuum conditions at room temperature. Overheating is prevented thanks to the efficient heat dissipation of the materials used, the suitable design and the low-power electronics on the board. A security mechanism was put in place to switch-off the device in case temperatures above $\rm35^{\circ}C$ are recorded. During operations, the baffle temperature increases only by a few degrees, and such high temperatures were never reached.

\section{Characterising stray light}
\label{section:Stray_light}
By placing photosensors around the IMC mirror, we can characterise the distribution of stray light and study the implications of its features and evolution. \correctionstwo{The dark-signal pedestal of each photodiode is determined from the mean signal measured during periods without light in the IMC cavity. The pedestals range from 4.60 to 32.21~$\mu$W, with an average value of 16.9~$\mu$W. The temporal fluctuations of the dark signal are evaluated separately for each photodiode with respect to its pedestal. The resulting RMS ranges from 0.01 to 0.77~$\mu$W, with an average value of 0.20~$\mu$W. These values are comparable to the 0.05--0.74~$\mu$W range reported shortly after installation~\cite{Ballester2022}, suggesting that the electronic performance remained stable.
For noise suppression, the noise level is defined as the $2\sigma$ threshold above the corresponding dark-signal pedestal, consistently with the procedure described in Ref.~\cite{Ballester2022}. Variations below this threshold are considered compatible with noise, while signals exceeding it are considered detectable. The resulting noise level ranges from 4.63 to 32.37~$\mu$W, with an average value of 17.31~$\mu$W. Thus, the sensitivity is channel-dependent, and a signal above approximately 32.4~$\mu$W is required to exceed the noise threshold of all photodiodes.
The signal-to-noise ratio is evaluated for each sensor by comparing the noise-suppressed signal with the corresponding noise level. It reaches values of up to approximately 20, with the highest values corresponding to the sensors contributing most strongly to the measured stray-light signal.
}

\subsection{Spatial distribution}
\label{subsection:Stay_light_spatial}
\begin{figure}[htb]
    \centering
    \includegraphics[width=\linewidth]{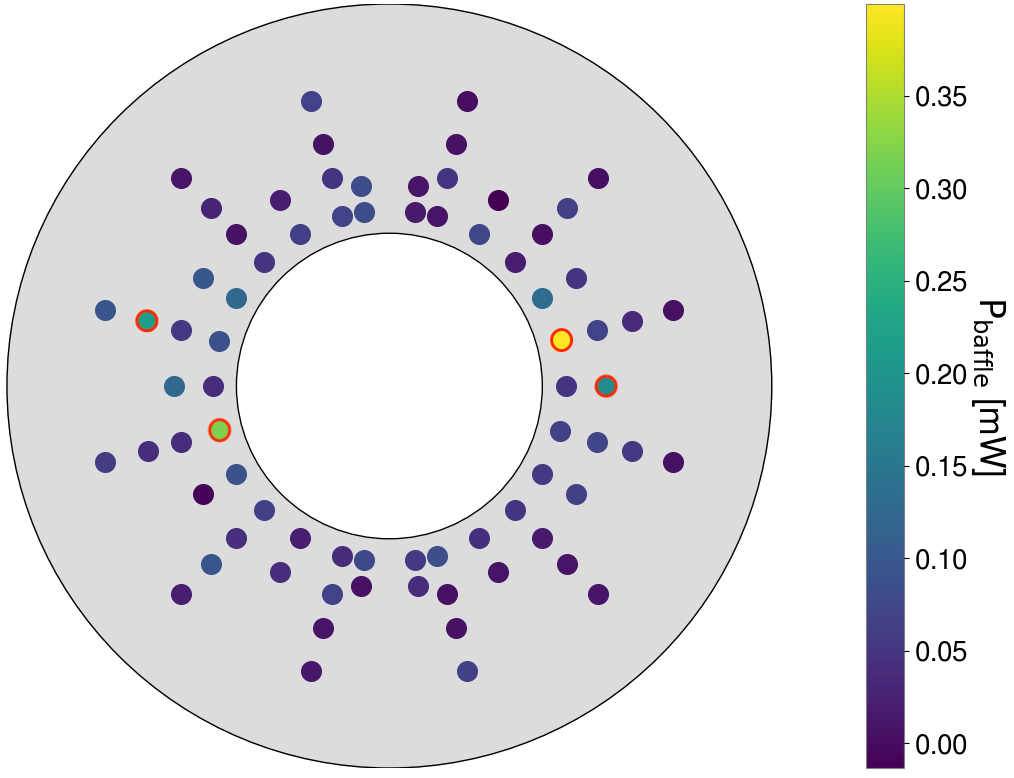}
    \caption{Two-dimensional diagram of the baffle indicating the average power measured by each sensor over October 2025. \corrections{The sensors circled in red are the main contributors to the signal measured by the instrumented baffle in October 2025.}}
    \label{fig:MeanPower2D}
\end{figure}

\begin{figure}[htb]
    \centering
    \includegraphics[width=\linewidth]{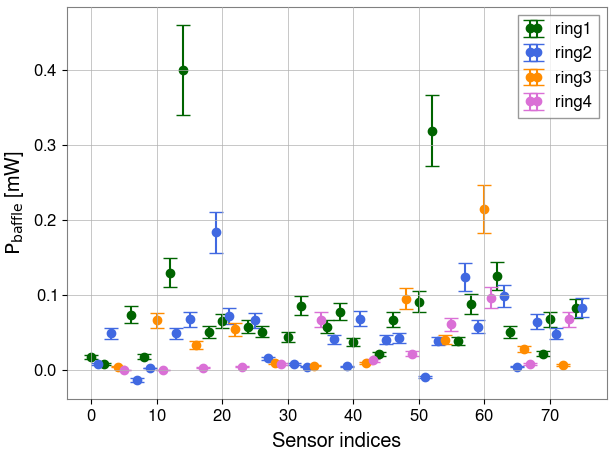}
    \caption{Power measured by the sensors in the four rings of the baffle and averaged over one month of stable operation in October 2025. The error bars \corrections{show the \correctionstwo{variability} of the stray light measurements over a month, corresponding to one standard deviation.}}
    \label{fig:MeanStdPower1D}
\end{figure}

We study the spatial distribution of stray light by comparing the power measured by the $76$ photodiodes of the baffle. The average total power measured by the baffle adds up to a few $\rm mW$, depending on the laser power.
Figures~\ref{fig:MeanPower2D} and \ref{fig:MeanStdPower1D} show the average power measured by each of the sensors for a one-month period of stable operation. We observe that the total power is not homogeneously distributed amongst the sensors. In particular, the zones that are further away from the mirror tend to see less scattered light. This observation complies with the expectations from simulations, which forecast stray light to be most important at low angles ~\cite{RomeroRodriguez2021}. Additionally, a small subset of sensors accounts for the majority of the total stray light contribution. These sensors can be considered the main contributors to the measured power, while the others constitute an isotropic stray light contribution below $\rm 0.1~ mW$.

The stability of the sensor readings agrees with the post-installation analysis and the initial instrumented baffle design described in Ref.~\cite{Ballester2022}. 
We observe that the main axis of symmetry of the stray light distribution is not normal to the incident beam but tilted by $\rm 15~deg.$ as displayed in Figure~\ref{fig:MeanPower2D}. This is coherent with the \corrections{initial} analysis of the data in 2021~\cite{Ballester2022}.  
As discussed in Ref.~\cite{Ballester2022}, this effect was explained through an analysis of the mirror maps of the IMC triangular cavity. Simulations of the light distribution inside the cavity, with and without the inclusion of the mirror maps, revealed that the dihedron mirrors, in particular the right one, were responsible for the peaks observed in sensors 14 and 52 (tilted by $\rm 15~deg.$ \corrections{in the anti-clockwise direction} with respect to sensors 18 and 56, which are aligned normal to the vertical axis). We study the evolution of this asymmetry in Section~\ref{section:Connection_ITF}.
Overall, the spatial features of stray light can be understood through a close comparison of observations with available simulations. Similarly to  Ref.~\cite{RomeroRodriguez2021} which demonstrates how different types of misalignment produce distinct stray light signatures, this approach opens the possibility of reconstructing deformations and aberrations of the IMC cavity mirrors. Incidentally, preliminary analyses for AdV+ Phase~II instrumented baffles suggest that the stray light patterns observed could be "reverse-engineered" to infer mirror aberrations such as point absorbers \cite{Macquet_2023,Andres_2025}.

\begin{figure*}[htb]
    \centering
    \includegraphics[width=\textwidth]{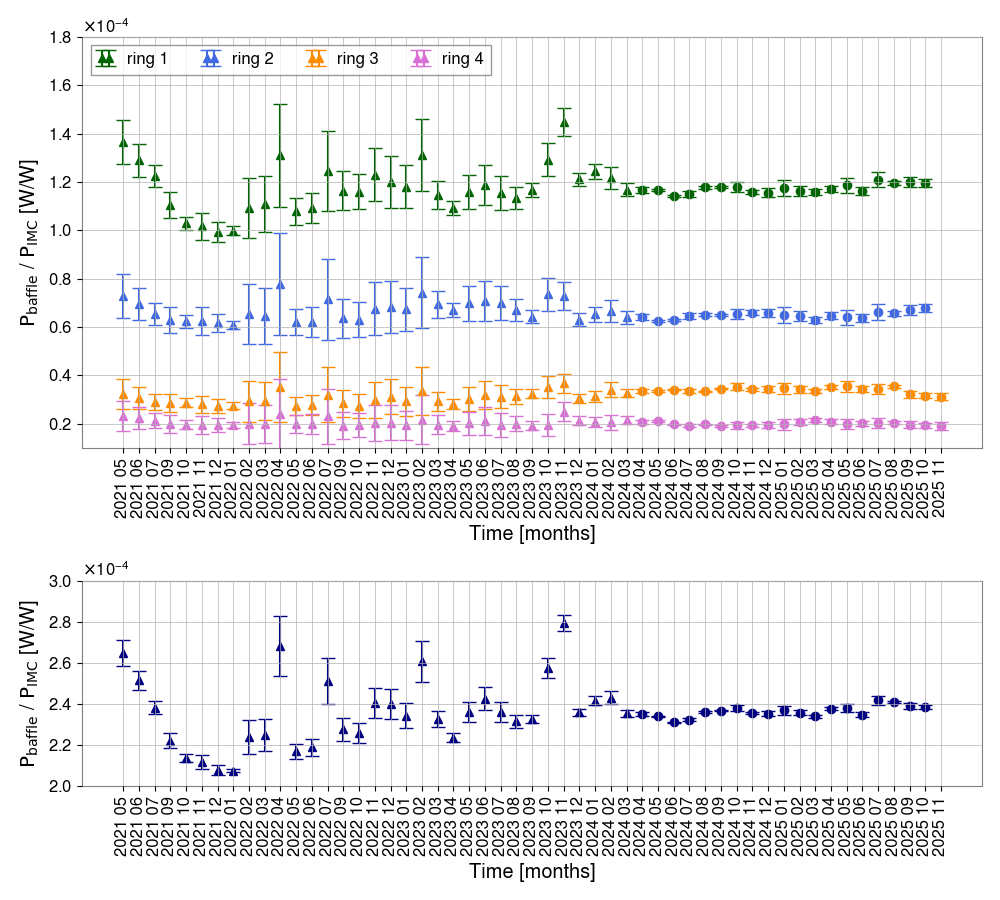}
    \caption{
    Power measured by the baffle, normalised by the IMC cavity power and averaged monthly between May 2021 and November 2025. August 2021 and 2022 are excluded due to a lack of data. \corrections{The triangular markers correspond to the commissioning period, while the circular markers show the science run period. Error bars represent the \correctionstwo{variability} of the stray light measurements, corresponding to one standard deviation.} The top panel shows measurements for each ring individually, while the bottom panel shows the total power.
    }
    \label{fig:Evolution}
\end{figure*}

\subsection{Temporal evolution}
\label{subsection:Stay_light_temporal}
After more than four years of continuous operation, spanning the O4 commissioning period and the O4b run, we assess the evolution of the scattered light distribution. Accounting for important commissioning activities occurring simultaneously with data taking between May 2021 and March 2024, we expect it to remain stable within a few standard deviations. After AdV+ joined O4 in April 2024, the stability is expected to improve substantially, making our results more relevant during that period.

We expect the total power measured by photosensors, $\rm P_{baffle}$, to follow the power variations in the IMC cavity $\rm P_{IMC}$. To allow for an objective comparison between May 2021 and November 2025, we normalise $\rm P_{baffle}$ by monthly-averaged $\rm P_{IMC}$ values\footnote{The channel \texttt{V1:INJ\_IMC\_TRA} of Virgo, \corrections{monitoring the power circulating in the IMC cavity}, was used.}. 
Results in Figure~\ref{fig:Evolution} show that rings are more illuminated by stray light when radially close to the beam, confirming that stray light is persistently more important at low angles \cite{RomeroRodriguez2021}. \corrections{By computing the ratio $\rm P_{baffle} / P_{IMC}$, we find that between 0.02\% and 0.03\% of the power circulating in the IMC cavity is seen by the photodiodes, with approximately 50\% of this power measured by the innermost ring. Since the active area of the photodiodes covers only about 1.2\% of the baffle surface, this value does not represent the total power incident on the baffle. Assuming that the scattered-light irradiance is uniformly distributed over the baffle surface, the total power intercepted by the baffle is estimated to be between 1.7\% and 2.5\% of the circulating cavity power.} \correctionstwo{Since the measured scattered-light distribution is not uniform, we further divide the baffle surface into concentric bands and account for the measured irradiance in each band. With this spatially resolved approach, we estimate that 1.4\% of the circulating power is incident on the baffle surface.}
For comparison, optical simulations presented in Ref.~\cite{RomeroRodriguez2021} predict significantly smaller fractions of the circulating power reaching the IMC baffle, with a maximum of $1.6 \cdot 10^{-3}\%$ under nominal optical configurations and $1.43 \cdot 10^{-3}\%$ for a cavity with a $10~\mu$rad mirror misalignment. \correctionstwo{While the simulations reproduce the main spatial features of the scattered-light distribution, the absolute level measured here is significantly larger. This difference may arise from the combination of non-ideal experimental conditions and scattering processes not fully accounted for in the simulations, which can affect the overall scattered-light level without necessarily modifying its main spatial distribution.}

\corrections{If one considers the status of AdV+ commissioning and operation, as shown in Figure~\ref{fig:VirgoStatusTimeline}, there is a noticeable shift in behaviour in early 2024. Indeed, Figure~\ref{fig:Evolution} evidences that, during the science run (April 2024 - November 2025), the mean value of the stray light measurements, i.e. the ratios between the power measured as stray light by the instrumented baffle to the cavity power, becomes essentially constant while their \correctionstwo{variability}, shown by the error bars, decreases significantly. The \correctionstwo{variability} of the stray light measurements has different physical origins related to the ITF status, in particular the IMC's. It demonstrates that the baffle can differentiate between the increased stray light resulting from commissioning activities from the stable stray light level characteristic of the science run.} 

The following section provides tentative explanations for the variations of $\rm \frac{P_{ baffle}}{P_{IMC}}$ on both short and long timescales.
\corrections{However, we want to emphasise that the IMC instrumented baffle is a demonstrator that such a device does not affect the performance of the cavity, as discussed in this section through the comparison of the baffle performance over time, and with the noise level evaluation in Section~\ref{section:Noise_level}. The IMC was chosen for this risk assessment because its impact on the overall ITF performance is expected to be limited. Unfortunately, the analysis of the IMC baffle data has a similarly limited reward, contrary to the instrumental baffles placed in the main arms \cite{Macquet_2023,Andres_2025}, hence limiting the scope and robustness of Section~\ref{section:Connection_ITF}.}

\section{Connection with the interferometer}
\label{section:Connection_ITF}

In this section, we evaluate the instrumented baffle in its role as an active monitoring device, examining the information it provides about the status of the interferometer and its subsystems. In particular, we focus on the injection system, laser beam misalignment, environmental perturbations, and transients. \corrections{Note that all results are tentative and exploratory lines of investigation anticipating the operation of instrumented baffles in the main arms.}

\subsection{Stability of the injection system}
\label{subsection:ITF_INJsys}
\begin{figure}
    \centering
    \includegraphics[width=\linewidth]{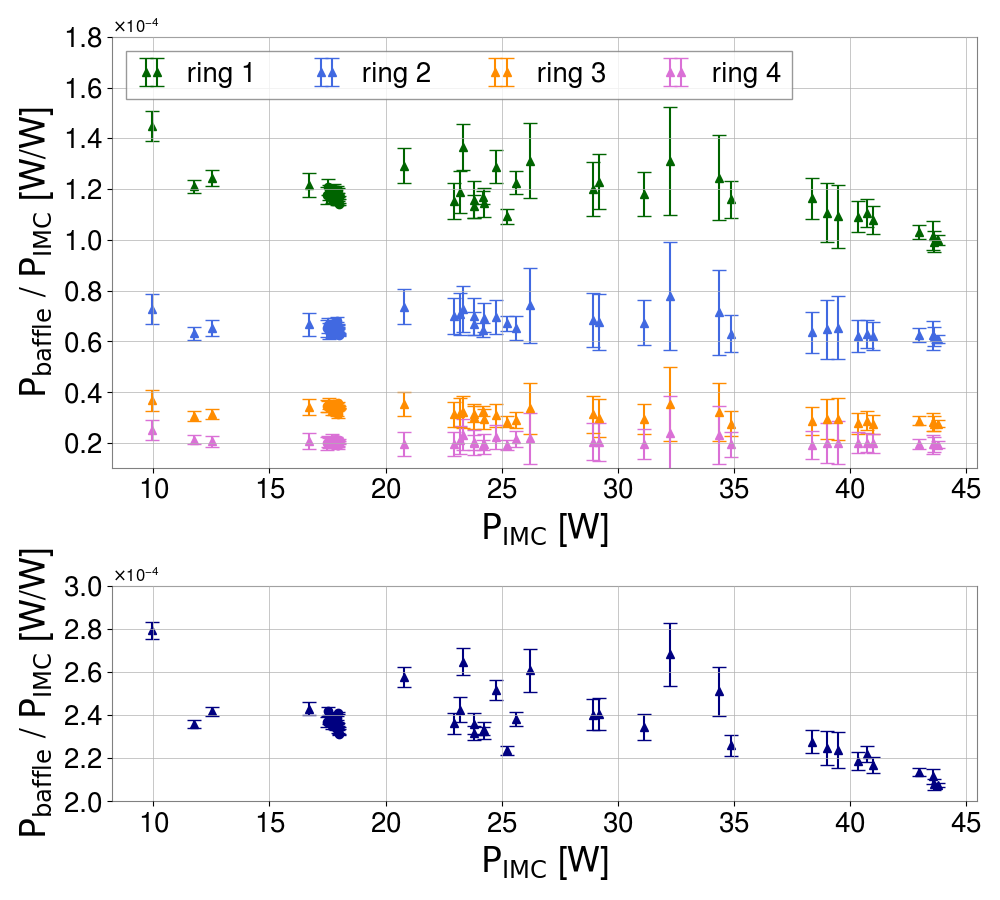}
    \caption{
    Monthly-averaged power normalised to the IMC cavity power as a function of the IMC power for each ring (\textit{top}) and the sum of the rings (\textit{bottom}).  August 2021 and 2022 are excluded due to a lack of data. \corrections{The triangular markers correspond to the commissioning period. The circular markers, all concentrated around the nominal O4 power $\rm P_{\rm IMC} \approx 17~W$, show the science run period. The error bars show the \correctionstwo{variability} of the stray light measurements over a month, corresponding to one standard deviation.}
    }
    \label{fig:Pratio_vs_Pimc}
\end{figure}

The injection system is closely related to the distribution of stray light, \corrections{especially because the IMC is an integral part of that system.}
The results in section \ref{subsection:Stay_light_temporal} were normalised by the laser power circulated in the IMC. Figure~\ref{fig:Pratio_vs_Pimc} presents the dependence of this ratio with $\rm P_{\rm IMC}$, separately for each baffle ring, where the size of the error bars is a direct indication of the stability of the interferometer in that period, with small uncertainties corresponding to periods of more stable interferometer operations.
In the past, it was measured that an increase in laser power can lead to the presence of higher-order modes and result in an increased variability of $\rm P_{ baffle} / P_{\rm IMC}$~\cite{Ballester2022}. Figure~\ref{fig:Pratio_vs_Pimc} shows that the ratio remains roughly constant and stable when varying the intra-cavity power, indicating that an increase in power does not imply a variation of the relative amount of stray light at large angles.  However, a trend is observed for the innermost ring, where most of the stray light is recorded: a linear decrease with $P_{\rm IMC}$ is observed in the region $\rm P_{\rm IMC}>35~W$. This can be attributed to different origins, including mirror-related thermal effects modifying the distribution of stray light in the cavity. The IMC mirrors are not equipped with a thermal compensation system as is the case for the main mirrors in the interferometer arms.  

In the course of the analysis, the correlation between the baffle measurements and Virgo auxiliary channels\footnote{The following Virgo channels was explored: \texttt{V1:INJ\_FMOD}, \corrections{related to the injection frequency modulation system;}
\texttt{V1:PSL\_PMC\_REFL}, \corrections{measuring the power in the Pre Mode Cleaner reflection port; and}
\texttt{V1:PSTAB\_PDa/b\_AC\_MONIT}, \corrections{monitoring the laser power stabilization photodiodes;}}, with the potential to spot changes in the stray light inside the cavity, were explored without decisive results. \corrections{A more thorough investigation with additional channels and metrics is left for future work.}

\corrections{\subsection{Beam alignment}}
\label{subsection:ITF_alignement} 
\begin{figure*}[htb]
    \centering

    \begin{subfigure}{\linewidth}
        \centering
        \includegraphics[width=\linewidth]{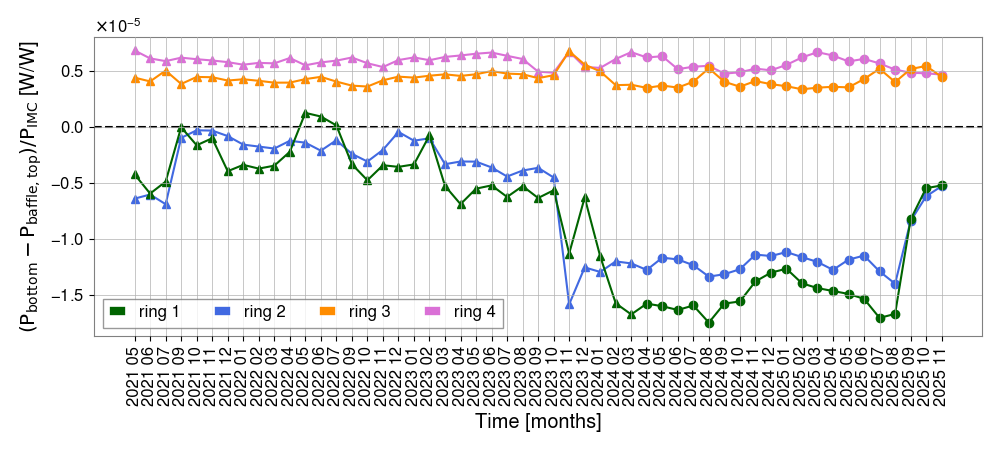}
    \end{subfigure}

    \vspace{0.5cm}
    
    \begin{subfigure}{\linewidth}
        \centering
        \includegraphics[width=\linewidth]{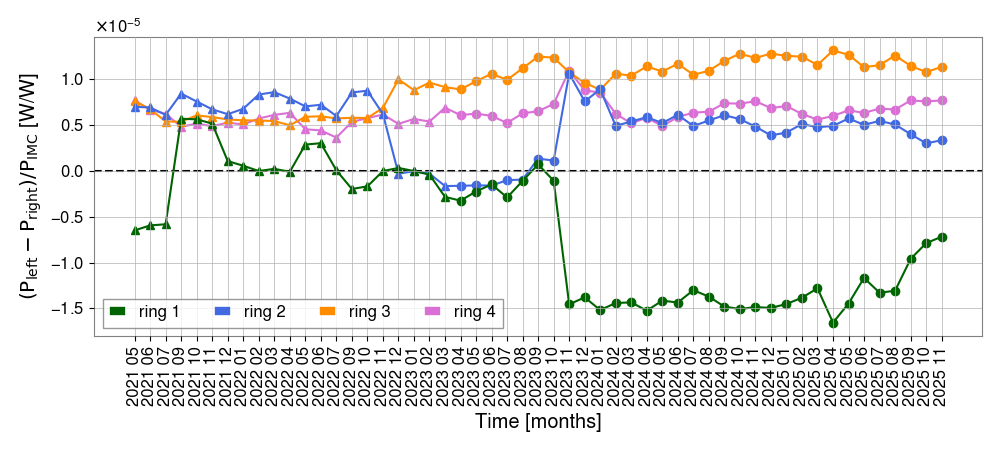}
    \end{subfigure}

    \caption{
        Evolution of the vertical (\textit{top}) and horizontal (\textit{bottom}) asymmetries of the scattered-light distribution, normalised to the intra-cavity power, over four years of IMC baffle operation. The differences $\mathrm{P_{bottom}-P_{top}}$ and $\mathrm{P_{left}-P_{right}}$ are computed using the baffle design shown in Figure~\ref{fig:BaffleSchema}. \corrections{The triangular markers correspond to the commissioning period, while the circular markers show the science run period.}
    }
    \label{fig:SymmetryEvolution}
\end{figure*}

As already shown in Figure~\ref{fig:MeanPower2D}, a right–left asymmetry was observed in the stray light distribution, consistent with what was seen during the initial baffle operation~\cite{Ballester2022}, and explained by the configuration of the mirror maps. \corrections{Note that the 15 deg. anti-clockwise tilt does not significantly affect the \correctionstwo{horizontal} asymmetry study since the main contributors to the signal remain in the same \correctionstwo{half-baffles} irrespective of whether the tilt is taken into consideration.} \correctionstwo{Instead, for the vertical asymmetry, sensors 14 and 52 would lie on the symmetry axis.} Here, we examine any significant changes in the vertical and horizontal asymmetries that may coincide with adjustments to the alignment strategy \corrections{or other activities affecting the stray light pattern} during commissioning.
Figure~\ref{fig:SymmetryEvolution} shows the evolution of the stray light pattern asymmetry over the four years of operation. \corrections{The variations are most pronounced in the two innermost rings of the baffle, where stray light accumulates.}  

The observed variations directly reflect changing conditions over the course of our study, demonstrating the instrumented baffle’s ability to monitor these changes. \corrections{For instance, in October 2021, an increase in both vertical and horizontal asymmetry coincided with investigations into the presence of higher-order modes in the IMC. In fact, higher-order modes can result in wider-angle scattering and a less uniform distribution of the scattered light. Similarly, in May 2022, a reduction in asymmetry followed power adjustments and IMC working point tuning. Indeed, demodulation phase mismatches can introduce systematic alignment errors that disrupt stray light symmetry. In November 2022, finding good alignment was proved increasingly difficult because of the appearance of high order modes. As a result, the beam was displaced horizontally on the end mirror to mitigate the impact of these mode resonances. It correlates with the increase in symmetry observed in the two innermost rings that month. Eventually, in November 2023, a drop in horizontal asymmetry in ring 1 correlated with the first full interferometer lock commissioning, during which IMC beam repositioning and alignment instabilities caused repeated unlocking and required manual realignment. During the O4 science run, the asymmetry remains high but stable until September 2025 when numerous IMC unlocks occurred and important IMC instability was reported, leading to an intervention. Please refer to Ref.~\cite{AdV+_O4} for a more global description of the O4 run. These cases highlight that the beam alignment and stability of the injection system are, in fact, entangled. While this reverse-engineering approach becomes more challenging during intensive commissioning, it demonstrates the instrument’s value for monitoring stray light and aiding interferometer operations.}

\begin{figure*}[htb]
    \centering
    \includegraphics[width=\linewidth]{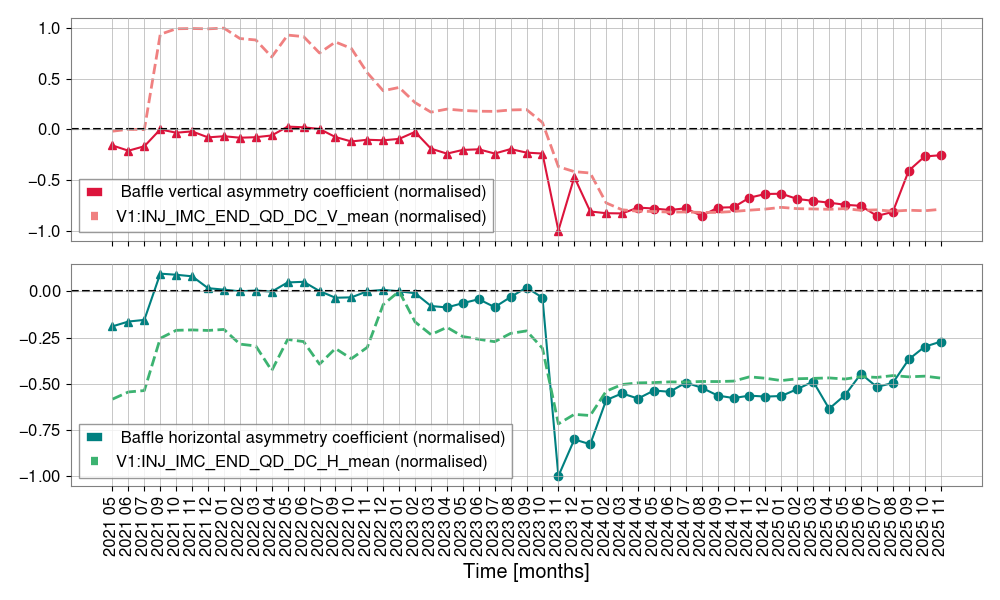}
    \caption{
    \corrections{Qualitative} comparison of vertical and horizontal \corrections{asymmetry} in the scattered-light distribution as measured by the \corrections{innermost ring} of IMC instrumented baffle (\corrections{respectively computed from the quantities $\mathrm{P_{bottom}-P_{top}}$ and $\mathrm{P_{left}-P_{right}}$, normalised by $\mathrm{P_{IMC}}$}) with the corresponding Virgo alignment channels (\corrections{\texttt{V1:INJ\_IMC\_END\_QD\_DC\_V\_mean} and \texttt{V1:INJ\_IMC\_END\_QD\_DC\_H\_mean}). The datasets were normalised to unity and the Virgo channels corrected for DC offset in order to facilitate the comparison. \corrections{For a quantitative description of the asymmetry, refer to Figure~\ref{fig:SymmetryEvolution}}}
    }
    \label{fig:Virgo_symmetry_evolution}
\end{figure*}

These variations can also be studied with Virgo auxiliary channels\footnote{The channels \texttt{V1:INJ\_IMC\_END\_QD\_DC\_V\_mean} and \texttt{V1:INJ\_IMC\_END\_QD\_DC\_H\_mean}, \corrections{monitoring the difference in power measured by the top-bottom and left-right} \correctionstwo{half-baffles} used for beam alignment in Virgo, respectively.}. \corrections{Figure~\ref{fig:Virgo_symmetry_evolution} shows that the asymmetry measured by the instrumented baffle follows the long-term evolution of the corresponding Virgo alignment channel, including the transition from predominantly near-zero to negative values in late 2023. The baffle symmetry coefficient indicates an important increase of asymmetry in favour of the top half of the baffle, which coincide with the manual realignment of the IMC discussed earlier. Although the month-to-month variability is significant, the consistently evolving variations indicates that the instrumented baffle is sensitive to changes in the IMC beam alignment.} 

\corrections{A more thorough investigation of \texttt{V1:INJ\_MC\_ty/tx} and \texttt{V1:BsX\_TX/TY/X/Y}, respectively monitoring the IMC alignment tilt and the beam pointing degrees of freedom is left for future work as the correlation of those channels did not provide conclusive results.}

\subsection{Environmental monitoring}
\label{subsection:ITF_ENV}
We anticipate that environmental perturbations—such as natural seismic activity, human-made disturbances, adverse weather, or significant temperature fluctuations—could lead to an increase in scattered light. We therefore performed a non-exhaustive study of the correlation between the baffle's stray light measurements and environmental Virgo channels. \corrections{The investigated quantities included both the mean evolution and the noise level, characterised through the RMS of the corresponding channels.}

\corrections{Among the auxiliary channels considered, particular attention was given to \texttt{V1:ENV\_MCB\_SEIS\_W\_rms}, which was identified during commissioning activities as one of the most plausible candidates for a coupling with the scattered-light level. We first investigated this channel as a representative test case for environmental noise coupling. Although sporadic common variations between the seismic noise level and the baffle power were observed when comparing monthly averaged quantities, no consistent correlation or causal relationship could be established over the studied period. Furthermore, local interferometer control systems are expected to suppress slow environmental variations, suggesting that any residual coupling would more likely appear through fluctuations at shorter timescales. To investigate this possibility, we also compared the short-timescale fluctuations of the baffle signal with the seismic noise level. The cross-correlation function presents a maximum at zero lag, with a broad triangular shape, indicating that some common temporal features are present in the two signals. However, the coherence remains close to zero below 0.5~Hz, showing that these similarities are not associated with a significant frequency-dependent coupling at low frequencies. The transfer function exhibits isolated features, with the largest amplitude observed around 0.2~Hz, but these features are not supported by a coherent phase response and therefore do not provide evidence of a stable physical coupling between seismic noise and the measured stray-light level.}

\corrections{The absence of a clear signature in the most likely candidate channel indicates that no dominant seismic contribution is observed in the current dataset. Nevertheless, we consider this result valuable, as it provides guidance for future environmental coupling studies. A more systematic investigation, including a broader set of auxiliary channels and frequency ranges, will be particularly relevant for future instrumented baffles in the main interferometer arms, where they can serve as dedicated monitors of environmental perturbations.}

\corrections{
Additionally, a strong daily periodicity was observed for sensors 9 and 11 in July 2024. We hypothesised that these sensors may be particularly sensitive to temperature variations. Further investigations into seasonal dependence, spatial correlations with neighbouring sensors, and localised temperature variations were conducted but proved inconclusive. While the periodicity matches that of the temperature sensor closest to the two sensors, no such behaviour is observed for other nearby sensors.}
\corrections{Finally, the building and mean baffle temperature measurements exhibit modulation with a 24-hour period. However, while the building temperature varies by approximately 1.5$^\circ$C, the baffle temperature remains much more stable, with a variation amplitude below 0.5$^\circ$C, and follows the environmental temperature variations with a delay of a few hours. This suggests that, although the baffle is efficiently thermally stabilised, it is not completely isolated from external temperature fluctuations. In addition, the median baffle temperature exhibits a slight drift over the one-week observation period. Therefore, while no direct correlation is observed, a possible contribution of slow thermal drifts to environmental coupling cannot be completely excluded.
}

\begin{figure}
    \centering
    \includegraphics[width=\linewidth]{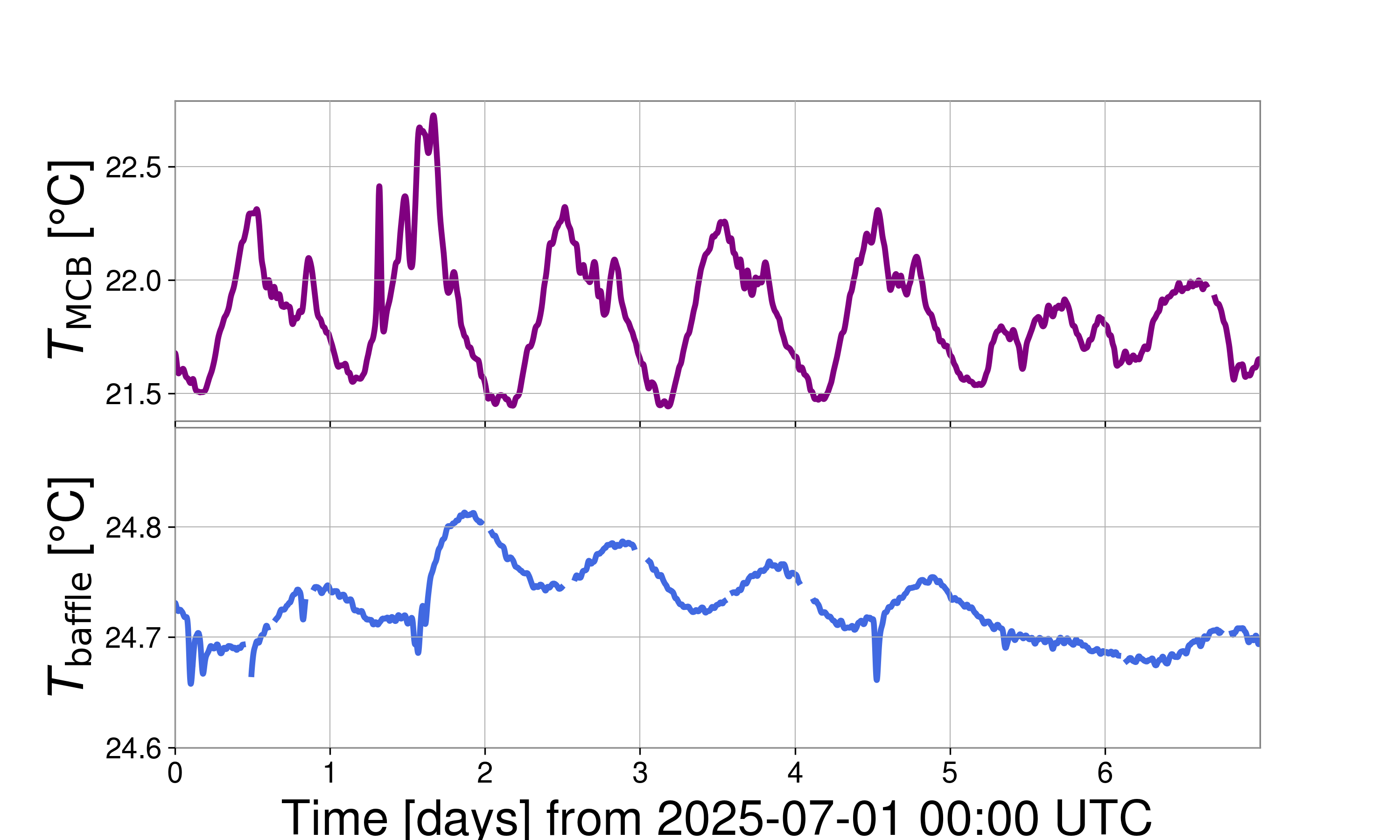}
    \caption{
    Comparison of the mean baffle temperature ($\rm T_{baffle}$) with the MCB temperature \corrections{(\texttt{V1:ENV\_MCB\_TE})} over a week in July 2025.
    }
    \label{fig:temperature}
\end{figure}

\subsection{Transients}

\begin{figure}[htb]
    \centering
    \includegraphics[width=\linewidth]{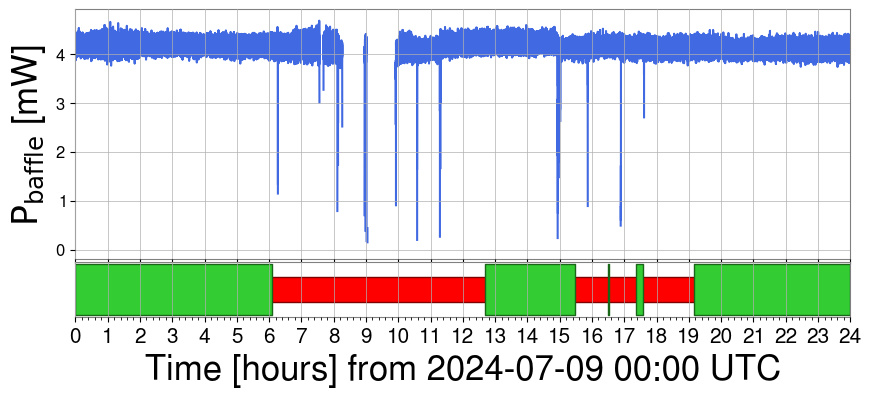}
    \caption{
    Comparison of the total baffle power, $\rm P_{baffle}$, with respect to interferometer lock status over 24 hours starting on July 9th, 2024, at 00:00 UTC. Red segments indicate unlocked periods, and green segments indicate locked periods.
    }
    \label{fig:glitches_characterisation}
\end{figure}

\begin{figure}[htb]
    \centering
    \includegraphics[width=\linewidth]{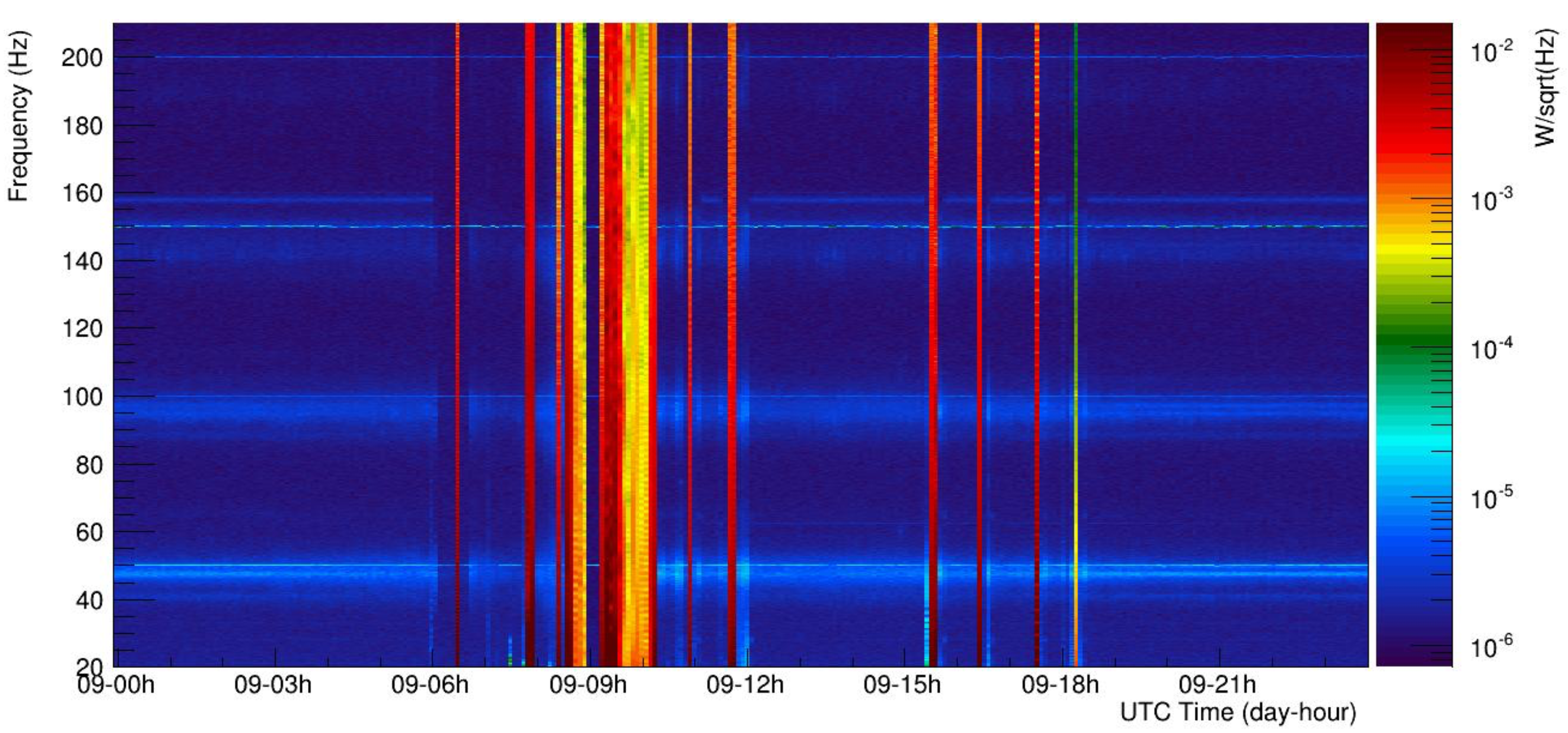}
    \caption{\correctionstwo{Time–frequency spectrogram of the IMC intra-cavity power, $\rm P_{IMC}$, over 24 hours starting on July 9th, 2024 at 00:00 UTC, shown over the frequency range [20, 210]~Hz. Colour represents the amplitude spectral density of $\rm P_{IMC}$, in W/$ \sqrt{\mathrm{Hz}}$, as indicated by the colour bar. Persistent horizontal features are visible near 50, 95, and 150 Hz, while strong, broadband transient events occur between 06h and 12h, and between 15h and 18h. This plot was made with the Virgo interferometer monitoring tool.}}
    \label{fig:spectrogram}
\end{figure}

The stray light distribution observed during stable operations can exhibit sudden changes when glitches and other transient signals occur, temporarily exceeding the steady stray light level. As shown in Figure~\ref{fig:glitches_characterisation}, these perturbations occur predominantly while the interferometer is unlocked. \correctionstwo{As a consequence, the photodiodes response is highly correlated, as strong transient events are simultaneously seen by all photodiodes. We use the IMC intracavity-power spectrograms, together with the interferometer lock status, to select two data segments on July 9th, 2024: a stable, locked period (00:00–06:00 UTC) and an unlocked period with a high rate of short broadband transients (06:00–12:00 UTC), marked by large, rapid variations in the $\rm P_{IMC}$ spectrograms (see Figure~\ref{fig:spectrogram}).  Note that Figures~\ref{fig:glitches_characterisation} and \ref{fig:spectrogram} are shown for the same time interval. For each segment, we calculate the mean pairwise correlation between sensors within each ring. Correlations are consistently higher during the transient-rich period: the ring-wise mean values range from 0.769 to 0.855, compared with 0.125 to 0.150 during the stable period. Averaged across the four rings, the mean correlation rises from 0.133 to 0.813. The corresponding medians are also higher in every ring, indicating that the increase is not driven solely by a small number of highly correlated sensor pairs. This consistent increase across all rings suggests that the transient-rich period involves a substantial common component in the sensor signals. Since the increase is also observed in rings beyond ring~1, it is not confined to the ring where steady scattered light is most concentrated. The increased correlation in the time segment with high transient rate support the interpretation that transients contribute to the sensors common behaviour.} 


\section{Noise level assessment}
\label{section:Noise_level}
Noise-hunting campaigns conducted prior to the start of observing runs are essential for identifying and mitigating potential sources of transient or persistent noise that could affect detector sensitivity. As part of the noise-hunting campaign preceding O4b, we assessed the noise level associated with the instrumented baffle. The effect of the baffle installation had previously been evaluated in 2021 by comparing key characteristics of the IMC cavity -such as losses, output power, finesse, and mismatch~\cite{Andres2023, Ballester2022}— and it was found that the baffle did not perturb measurement conditions.

\begin{figure*}[htb]
    \centering
    \includegraphics[width=\linewidth]{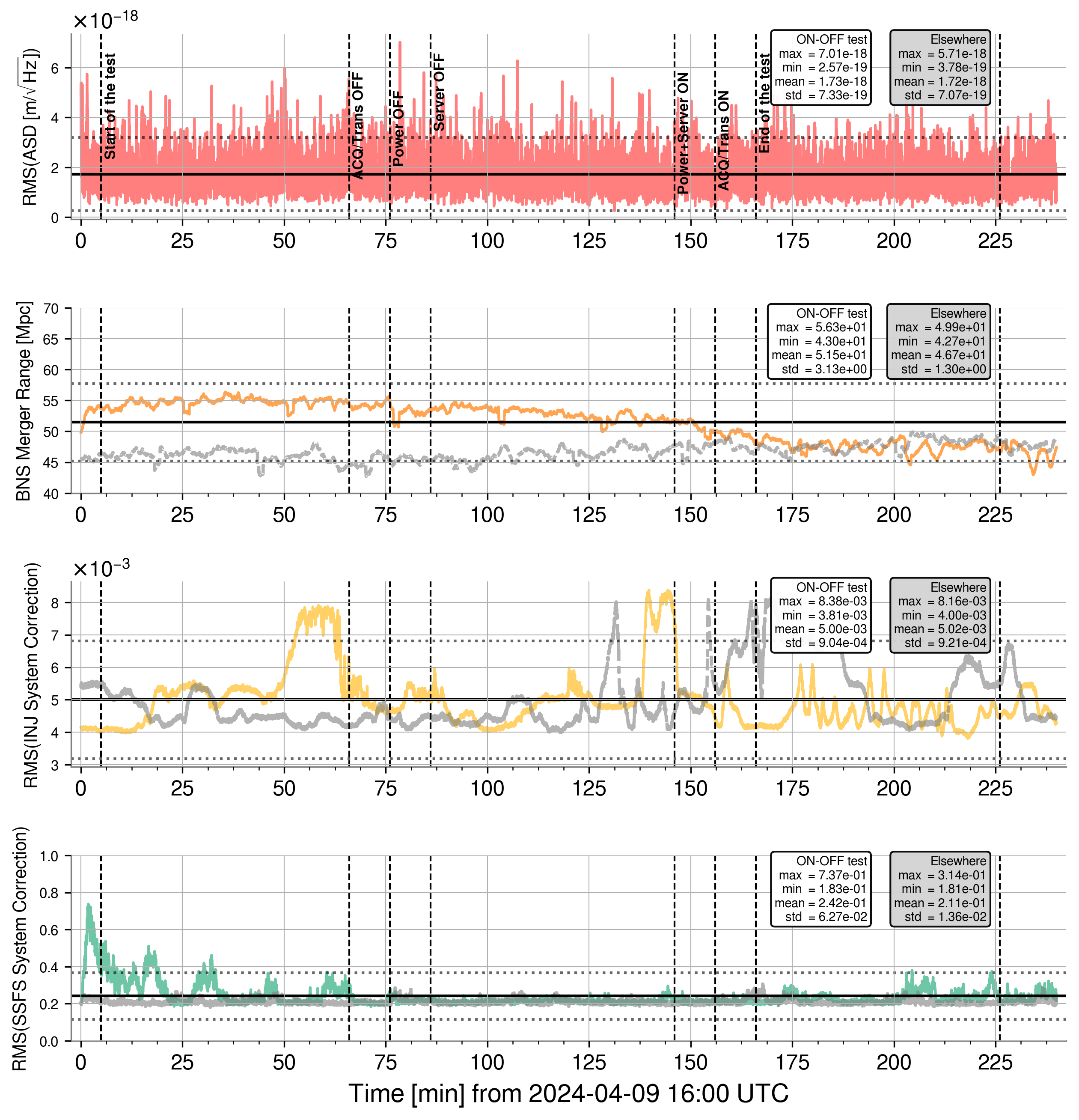}
    \caption{\corrections{The noise level assessment performed on March 9, 2024, showing the RMS of the calibrated strain Amplitude Spectral Density (ASD), the binary neutron star (BNS) inspiral range, and the RMS of the correction signals of the injection (INJ) and Second Stage Frequency Stabilization (SSFS) control loops. The quantities are computed from the corresponding Virgo data channels: \texttt{V1:Hrec\_hoft\_16384Hz\_rms}, \texttt{V1:Hrec\_Range\_BNS}, \texttt{V1:INJ\_EOM\_CORR\_rms}, and \texttt{V1:SSFS\_corr\_50Hz\_rms}, respectively. The coloured curves correspond to the measurements during the noise level assessment on March 9, 2024, while the gray curves show the same quantities evaluated at a reference time chosen outside the assessment period.}}
    \label{fig:MeanStdPower1Dbis}
\end{figure*}

The noise assessment was performed on March 9, 2024, and involved the sequential switch-off and switch-on of the instrumented baffle components. A stable dataset of one hour was collected, followed by 30 minutes of data acquisition after each successive step, which included switching off the data transmission system, the server, and the power supply, then turning them back on after another hour of data collection. This sequence of steps is shown in Figure~\ref{fig:MeanStdPower1Dbis} along with the RMS of some channels that are closely linked to the IMC behaviour. We include the amplitude spectral density (ASD) of the interferometer\corrections{, providing a sensitive indicator of any broadband or transient excess noise introduced by the baffle instrumentation}.  We also monitor the two correction signals related to the laser injection system (INJ) and laser frequency stabilisation (SSFS, i.e. the second step of the frequency stabilisation)\corrections{, as these feedback signals directly reflect changes in the control effort required to maintain the laser frequency and alignment, making them sensitive probes of any disturbance coupled to the interferometer through the baffle \cite{AdV+_O4}}. Additionally, we present the reachable distance for binary neutron star (BNS) merger detection (in Mpc)\corrections{, which provides a global figure of merit for detector sensitivity and allows us to verify that the intervention produced no measurable degradation in astrophysical performance}.

The intervention had no significant impact on interferometer operation. No disturbances in the RMS of the ASD were observed beyond typical temporal variations. Complementarily, the spectrogram of the ASD was obtained, for which we observe no disturbances either. Similarly, the spectrogram of the sensor in the IMC building, used to monitor magnetic interference, showed no changes, ruling out magnetic effects. We examined the error signals sent by the IMC to the laser during the first step of laser frequency stabilisation and observed no variations beyond the usual, nor any causal connection to the intervention. The structures observed in the INJ correction channel were also present in the SSFS correction signal, with no evidence of a causal connection with the \correctionstwo{interventions}. Moreover, a comparison with the reference period shows that these features are within the typical variation range of these channels.

\section{Conclusion}
An instrumented baffle was installed in the IMC cavity as part of the AdV+ Phase I upgrade. The four-year data analysis presented here aims to study the performance and stability of the instrument across successive periods of upgrades, commissioning, and data taking. We find that the instrumented baffle does not perturb the IMC or ITF operations. The detailed analysis shows that the stray light distribution matches expectations from simulations, with a strong radial asymmetry originating from the cavity layout.
We observe that the innermost ring accounts for 50\% of the stray light. The stray light measured by the baffle, normalised to the intra-cavity power, remains stable between \corrections{0.02\% and 0.03\% (estimated to 1.7\% and 2.5\% for the whole baffle surface)}.
Our results show that the baffle can be used as a stray light monitoring device close to the mirrors. The data analysis provides insight into the cavity stability, the presence of transient signals, and the beam alignment. The laser power increase foreseen in future AdV+ upgrades, aimed at enhancing sensitivity, further underscores the importance of monitoring these effects.
The IMC instrumented baffle serves as a proof of concept and a demonstrator of the applications of this technology, which is anticipated to play a key role in future-generation gravitational-wave interferometers, including the Einstein Telescope. Instrumented baffles with more sensors and faster readout \corrections{were recently installed} in the main arms of the Virgo detector, as part of the AdV+ Phase~II upgrades, to monitor the light in the main cavities.

\section*{Acknowledgments}
The authors would like to thank A. Lundgren and M. Gosselin for helpful discussions. The authors gratefully acknowledge the European Gravitational Observatory and the Virgo Collaboration for providing access to the data and facilities. 

This work is partially supported by the Spanish MCIN/AEI/10.13039/501100011033 under the Grants No.~PID2020-113701GB-I00, PID2023-146517NB-I00, and CEX2024-001441-S, some of which include ERDF funds from the European Union, and by the MICINN with funding from the European Union NextGenerationEU (PRTR-C17.I1) and by the Generalitat de Catalunya. IFAE is partially funded by the CERCA program of the Generalitat de Catalunya.
This project has received funding from the European Union’s Horizon Europe research and innovation programme under the Marie Skłodowska-Curie grant agreement No.~10181337.

\section*{Data availability statement}
The data supporting the findings of this study consist of Virgo auxiliary detector channels and measurements from the instrumented baffle. Owing to the Virgo Collaboration data access policy, these data are not publicly available. Processed data and analysis products supporting the results presented in this work can be made available by the corresponding author upon reasonable request, subject to the Virgo Collaboration's data access and approval procedures.

\section*{Statement of Provenance}
This is the Accepted Manuscript version of an article accepted for publication in Classical and Quantum Gravity. IOP Publishing Ltd is not responsible for any errors or omissions in this version of the manuscript or any version derived from it. The Version of Record is available online at  https://doi.org/10.1088/1361-6382/aeb21f.

\section*{References}
\bibliography{references}

@article{Andres2023,
title={Instrumented baffle for the Advanced Virgo input mode cleaner end mirror},
volume={107},
doi={10.1103/physrevd.107.062001},
url={https://doi.org/10.1103/physrevd.107.062001},
number={6},
journal={PRD},
author={Andrés-Carcasona, M. and others},
year={2023},
}

@article{Ballester2022,
title={Measurement of the stray light in the Advanced Virgo input mode cleaner cavity using an instrumented baffle},
volume={39},
doi={10.1088/1361-6382/ac6a9d},
url={https://doi.org/10.1088/1361-6382/ac6a9d},
number={11},
journal={CQG},
author={Ballester, O and others},
year={2022},
pages={115011},
}

@article{RomeroRodriguez2021,
doi={10.1088/1361-6382/abce6b},
url={https://doi.org/10.1088/1361-6382/abce6b},
year={2020},
volume={38},
number={4},
pages={045002},
author={A Romero-Rodríguez and others},
title={Determination of the light exposure on the photodiodes of a new instrumented baffle for the Virgo input mode cleaner end-mirror},
journal={CQG},
}

@article{Accadia2012,
doi={10.1088/1748-0221/7/03/P03012},
url={https://doi.org/10.1088/1748-0221/7/03/P03012},
year={2012},
volume={7},
number={03},
pages={P03012},
author={{Virgo Collaboration}},
title={Virgo: A laser interferometer to detect gravitational waves},
journal={JINST},
}

@article{Acernese2014,
title={Advanced Virgo: A second-generation interferometric gravitational wave detector},
volume={32},
doi={10.1088/0264-9381/32/2/024001},
url={https://doi.org/10.1088/0264-9381/32/2/024001},
number={2},
journal={CQG},
author={{Virgo Collaboration}},
year={2014},
pages={024001},
}

@article{Acernese2019,
title={Increasing the Astrophysical Reach of the Advanced Virgo Detector via the Application of Squeezed Vacuum States of Light},
author={{Virgo Collaboration}},
journal={PRL},
volume={123},
issue={23},
pages={231108},
numpages={10},
year={2019},
doi={10.1103/PhysRevLett.123.231108},
url={https://doi.org/10.1103/PhysRevLett.123.231108},
}

@article{Acernese2023_future,
doi={10.1088/1742-6596/2429/1/012040},
url={https://doi.org/10.1088/1742-6596/2429/1/012040},
year={2023},
volume={2429},
number={1},
pages={012040},
author={{Virgo Collaboration}},
title={Advanced Virgo Plus: Future Perspectives},
journal={JPCS},
}

@article{Acernese2023_status,
doi={10.1088/1742-6596/2429/1/012039},
url={https://doi.org/10.1088/1742-6596/2429/1/012039},
year={2023},
volume={2429},
number={1},
pages={012039},
author={{Virgo Collaboration}},
title={The Advanced Virgo+ status},
journal={JPCS},
}

@article{Macquet_2023,
title={Simulations of light distribution on new instrumented baffles surrounding Virgo end mirrors},
volume={40},
doi={10.1088/1361-6382/acc166},
url={https://doi.org/10.1088/1361-6382/acc166},
number={7},
journal={CQG},
author={Macquet, A and others},
year={2023},
pages={077001},
}

@article{Andres_2025,
title={Performance of an instrumented baffle placed at the entrance of Virgo's end mirror vacuum tower during O5},
author={Andr\'es-Carcasona, M. and others},
journal={Phys. Rev. D},
volume={111},
issue={4},
pages={042001},
numpages={9},
year={2025},
month={Feb},
publisher={American Physical Society},
doi={10.1103/PhysRevD.111.042001},
url={https://doi.org/10.1103/PhysRevD.111.042001},
}

@misc{AdV+_O4,
title={Advanced Virgo during the LIGO-Virgo-KAGRA fourth observing run},
author={{Virgo Collaboration}},
year={2026},
eprint={2607.26872},
archivePrefix={arXiv},
primaryClass={gr-qc},
url={https://arxiv.org/abs/2607.26872},
}

\end{document}